# nature COMMUNICATIONS

ARTICLE

Received 24 Apr 2015 | Accepted 31 Aug 2015 | Published 6 Oct 2015

DOI: 10.1038/ncomms9558

OPEN# Effect of interfaces on the nearby Brownian motion

Kai Huang[1] & Izabela Szlufarska[1,2]Near-boundary Brownian motion is a classic hydrodynamic problem of great importance in a variety of fields, from biophysics to micro-/nanofluidics. However, owing to challenges in experimental measurements of near-boundary dynamics, the effect of interfaces on Brownian motion has remained elusive. Here we report a computational study of this effect using μs-long large-scale molecular dynamics simulations and our newly developed Green–Kubo relation for friction at the liquid–solid interface. Our computer experiment unambiguously reveals that the $t^{-3/2}$ long-time decay of the velocity autocorrelation function of a Brownian particle in bulk liquid is replaced by a $t^{-5/2}$ decay near a boundary. We discover a general breakdown of traditional no-slip boundary condition at short time scales and we show that this breakdown has a profound impact on the near-boundary Brownian motion. Our results demonstrate the potential of Brownian-particle-based micro-/nanosonar to probe the local wettability of liquid–solid interfaces.[1] Materials Science Program, University of Wisconsin, Madison, Wisconsin 53706-1595, USA. [2] Department of Materials Science and Engineering, University of Wisconsin, 1509 University Avenue, Madison, Wisconsin 53706-1595, USA. Correspondence and requests for materials should be addressed to K.H. (email: huangk05@gmail.com) or to I.S. (email: szlufarska@wisc.edu).NATURE COMMUNICATIONS | 6:8558 | DOI: 10.1038/ncomms9558 | www.nature.com/naturecommunications      1© 2015 Macmillan Publishers Limited. All rights reserved.

It is now well known that the velocity autocorrelation function (VAF) of a Brownian particle in a bulk liquid does not decay exponentially as predicted by the Einstein–Ornstein–Uhlenbeck theory[1,2], but instead it follows a $t^{-3/2}$ algebraic decay at the hydrodynamic long-time limit[3]. Such a $t^{-3/2}$ long-time tail, first discovered for neat liquid in the seminal computer experiment by Alder and Wainwright[4], has been recently observed experimentally for single Brownian particle trapped in optical tweezers experiments[5–8]. When the optical trap is stiff enough, the power spectral density (PSD) of the particle's position exhibits a hydrodynamic resonance[6]. The slow VAF decay in time and concomitant resonance in PSD are related to each other by Fourier transform and are both the result of the hydrodynamics coupling between the Brownian particle and bulk liquid. This coupling is mediated by the vorticity generated by the particle. When the Brownian particle moves towards a boundary and when the particle-generated vortex reaches the boundary (see Fig. 1a), it is clear that the hydrodynamic coupling must change, but the question is how.

Since Lorentz[9], theorists have attempted to answer this question from hydrodynamic and lattice-Boltzmann calculations[10–12]. However, consensus has not been reached so far even for the simplest case in which a spherical particle is immersed in Newtonian liquid bounded by a flat surface with no-slip boundary condition (the slip length $L_s$ defined in Fig. 1b vanishes). For instance, Berg–Sørensen and Flyvbjerg[11] developed models for PSD, which when Fourier transformed predict a persistent $t^{-3/2}$ long-time tail in VAF near a boundary. This is the same functional form as in a bulk liquid, but with a reduced amplitude. On the other hand, Felderhof[12] derived a model that exhibits an algebraic decay of $t^{-5/2}$ (see Fig. 1c) for the same case of a near-boundary particle. Comparisons of these predictions to experimental measurements have only been made in recent years. For instance, Jeney et al.[13] measured the VAF of a near-boundary Brownian particle trapped with optical tweezers and reported a $t^{-5/2}$ long-time tail. However, a later optical tweezers experiment by Jannasch et al.[14] observed a reduced magnitude of PSD in the low-frequency (long-time tail) limit and a suppression of a hydrodynamic resonance near a boundary (see Fig. 1d), for which the phenomenon was explained using Flyvbjerg's model with the $t^{-3/2}$ long-time decay of VAF. Although contradicting each other, both experiments suffer from large statistical uncertainties in the long-time or low-frequency limit. Therefore, so far, experimental findings on the asymptotic behaviour of near-boundary Brownian particle have been inconclusive.

Molecular dynamics (MD) simulations provide a powerful tool for addressing the above questions, because they do not require a priori assumptions about molecular phenomena at the liquid–solid interface[3,15,16]. However, typical MD simulations of dynamic phenomena suffer from the limitations of accessible time scales. Here we overcome this challenge by using μs-long large-scale MD simulation and our recently developed Green–Kubo (GK) relation for liquid–solid friction[17]. Thanks to this new technique, our simulations are able to measure the boundary condition at the zero shear-rate limit, consistent with the regime explored in the optical tweezers experiments. We study the Brownian motion of a near-boundary nanoparticle immersed in approximately half a million solvent molecules. The use of such a large simulation system ensures that finite-size effects (for example, the effect of the acoustic wave travelling through the entire simulation domain within the time of the measurement) can be excluded. Unlike the optical tweezers experiments that trap the Brownian particle in all three dimensions, we only constrain the particle in the direction perpendicular to the surface by applying a harmonic potential. Therefore, the particle performs free Brownian motion parallel to the surface and its long-time tail in the lateral directions will not be truncated nor affected by parallel constraints, as it can happen in experiment[13]. By assigning strong interaction between solvent and solid atoms, we create a fully wetted surface that exhibits no slip in the long-time limit, which is consistent with typical experimental situations.

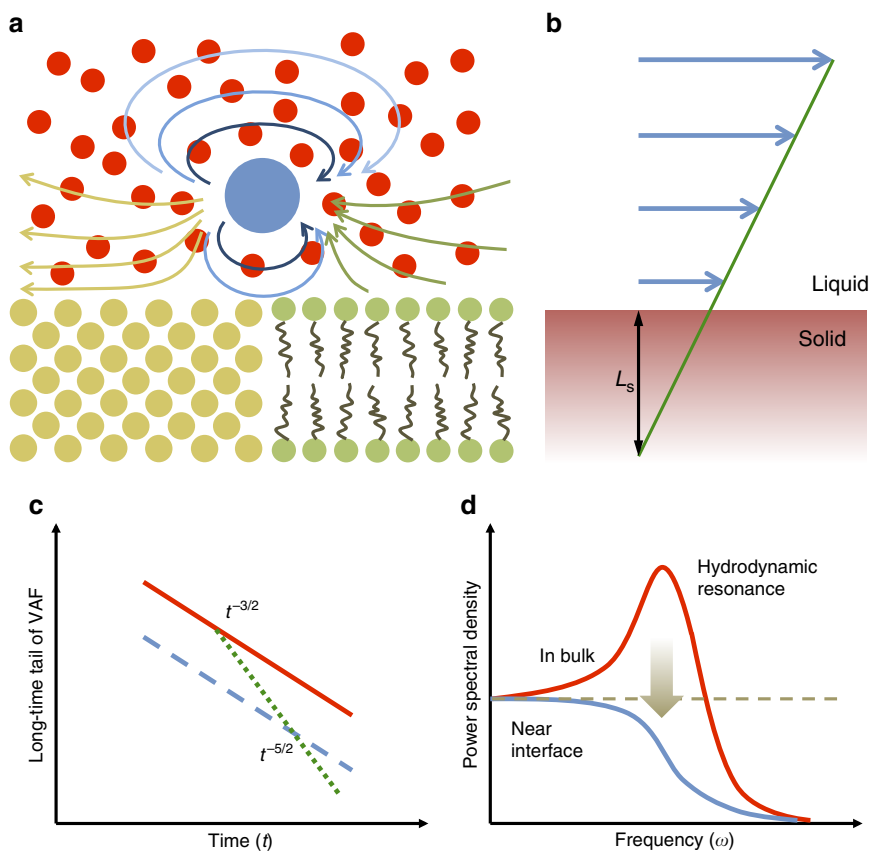

**Figure 1 | Schematic of near-boundary Brownian motion.** (**a**) The presence of the boundary affects the diffusivity of the Brownian particle and its hydrodynamic coupling to the solvent. This effect is expected to be surface-specific and understanding it is useful for advanced sensing applications. (**b**) Slip length $L_s$ defined to characterize the boundary condition of a partially wetted surface. (**c**) Log–log representation of the long-time tails. According to different theories, the $t^{-3/2}$ long-time tail in bulk liquid (solid red) could be persistent but suppressed (dashed blue) near a boundary, or transition into a $t^{-5/2}$ decay (dotted green). (**d**) The resonance peak due to the particle–solvent hydrodynamic coupling in bulk liquid is found to be suppressed as the particle approaches a boundary.

In the following sections we report results of our MD simulations of the effect of interface on the nearby Brownian motion. We demonstrate that boundary condition is inherently dynamic and that the VAF of a near-boundary Brownian particle is sensitive to the boundary relaxation. Such sensitivity can enable the use of Brownian particles to probe the properties of liquid–solid interfaces.

## Results

**Effect of interface on diffusivity.** Before studying the VAF of the Brownian motion, we first demonstrate that the hydrodynamic theories can quantitatively predict the diffusion coefficient for a motion of a nanoparticle in our simulations. Although the functional form of the near-boundary VAF of Brownian motion remains controversial, it is generally agreed that near a fully wetted solid surface (which yields a no-slip boundary condition), the diffusion of a Brownian particle is slower than in bulk. Such boundary confinement was first demonstrated by Lorentz[9], who predicted a reduced diffusion constant $D_w = D_b(1 - \frac{9R}{16h})$, where $R$ is the hydrodynamic radius of the particle[18] and $h$ is the distance between the centre of the particle and the solid wall. In this equation, $D_b = \frac{kT}{c\pi\eta R}$ is the Einstein–Stokes relation[1] for Brownian diffusion constant in bulk liquid, with $kT$ being the thermal energy and $\eta$ the viscosity of liquid. The constant $c$ is equal to 4 when there is no friction between the nanoparticle and the surrounding solvent[3,18] (as is the case in our simulations). In our simulations, we use mean square displacement to calculate the diffusion coefficient of a nanoparticle that is 125 times heavier than solvent and with $R = 3\sigma$, where $\sigma$ is the reduced unit of distance employed in simulation and corresponds to a few angstroms in real units. To ensure that we truly test the effect of boundary confinement, we constrain our nanoparticle's motion very close to the wall with $h = 5\sigma$. Diffusion coefficients (which are time integrals of VAF) calculated both in bulk and near a liquid–solid interface are shown in the inset of Fig. 2 and they are compared with the predictions of Lorentz and Einstein–Stokes theories, respectively. In general, a good agreement is reached between simulations and theory, which demonstrates that our simulations of near-boundary motion capture correctly the physics that is already well established. One should point out that the original Lorentz theory was developed assuming a point particle that is reasonably far away from the wall ($R/h \ll$). However, as shown in Fig. 2 and as further demonstrated in Supplementary Note 1, this theory can be successfully applied to the geometries considered in our simulations.

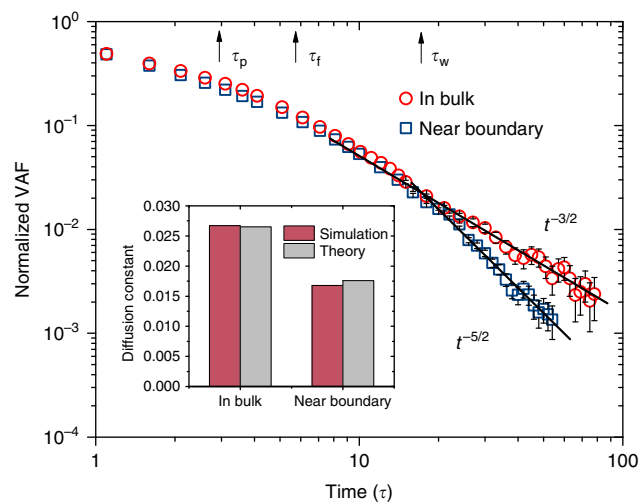

**Figure 2 | Diffusion constant and VAF of Brownian particle.** Log–log plot of the VAFs (normalized by its value at time zero) of a nanoparticle in bulk liquid (circles) and near a static no-slip boundary (squares). Solid black lines labelled $t^{-3/2}$ and $t^{-5/2}$ are added to guide the eye. In the inset, the diffusion coefficients of the nanoparticle in bulk and near boundary are compared with their theoretical predications at Brownian limit. $\tau_p$ is a characteristic timescale equal to $m_p/\gamma$, where $m_p$ is the mass of the particle and $\gamma$ is the Stokes friction coefficient. $\tau_f$ and $\tau_w$ are defined in the main text. Data points represent averages over 2,000 independent simulations for bulk VAF and 5,000 for near-boundary VAF. Error bars are obtained as s.d. these averages.

**Effect of interface on VAF.** We now focus on the highly debated VAF of the nanoparticle, as this function encodes the full dynamic information about the Brownian motion. We first discuss the results in bulk liquid, shown with red circles in Fig. 2. After time $\tau_f = R^2\rho/\eta = 5.7$ (where $\tau_f$ is the time scale of developing a vortex comparable to the size of the particle and $\rho$ is the liquid density), the VAF decays according to the functional form $A_b t^{-3/2}$ and it is the long-time tail behaviour. According to theory[19], the amplitude $A_b$ of the bulk $t^{-3/2}$ long-time tail depends only on the shear viscosity of the liquid

$$A_b = \frac{kT\rho^{1/2}}{12(\pi\eta)^{3/2}}. \quad (1)$$

By fitting the long-time tail measured in our simulations and using equation (1), we find the liquid shear viscosity of 1.0. This value is in a good agreement with $\eta = 1.1$ measured directly from simulations using the following GK relation[20]

$$\eta = \frac{V}{kT}\int_0^\infty \langle P_{xy}(t)P_{xy}(0)\rangle dt, \quad (2)$$

where $V$ is the volume of bulk liquid and $P_{xy}$ represents off-diagonal (shear) components of the stress tensor.

When the nanoparticle approaches a boundary, the VAF starts to deviate strongly from the bulk VAF after $\tau_w = h^2\rho/\eta = 15.8$, which is the time when particle-induced vorticity reaches the interface. As shown in Fig. 2, in the long-time limit, the VAF of the near-boundary nano- exhibits a $t^{-5/2}$ behaviour, instead of the $t^{-3/2}$ behaviour observed in the bulk. Our finding conclusively demonstrates a transition from the bulk $t^{-3/2}$ long-time tail (red solid line in Fig. 1c) to a $t^{-5/2}$ one (green dotted line in Fig. 1c) near a boundary and excludes the possibility of a persistent but reduced $t^{-3/2}$ long-time tail (blue dashed line in Fig. 1c), previously proposed in literature[11].

For the VAF to serve as a local probe of the nanoparticle's environment[6], it is important to understand not only the qualitative trend in VAF but also the impact of the environment on VAF's amplitude. The amplitude $A_w$ of the $t^{-5/2}$ long-time tail near a boundary has been recently derived by Felderhof[12] using hydrodynamic theories

$$A_w = CkT\frac{R^2}{\eta}\left(\frac{\rho}{4\pi\eta}\right)^{3/2}, \quad (3)$$

where $C = h^2/R^2 - 5/9 + 2\rho_p/9\rho + (1-\rho_p/\rho)R/8h$ with $\rho_p$ being the density of the Brownian particle. Unlike $A_b$, which depends only on the liquid properties, $A_w$ is also dependent on the density $\rho_p$ of the particle. To test the predictive power of equation (3), we keep the size of our nanoparticle constant and we increase its mass from 125 to 375 in reduced units (corresponding to solute-to-solvent density ratios of 1.6–4.8). In Fig. 3, we plot the VAFs of the Brownian particles with varying masses and in the inset we compare the amplitudes of the $t^{-5/2}$ long-time tails with the theoretical predictions (see Supplementary Figure 1 and Supplementary Table 1 for fitting details). Although a reasonably good match between simulation and theory is achieved for the most massive particle, a remarkable deviation is found for the lightest particle. Specifically, the numerical result in the case of small nanoparticle is significantly suppressed compared with the theoretical value. Such discrepancy suggests that the time-scale separation between the solute and solvent dynamics may not be sufficient for the lightest particle.

**Relaxation of boundary condition.** An extreme separation of time scales between the fast relaxation of solvent transport coefficients and the slow Brownian motion of a nanoparticle is a major assumption of hydrodynamic theories of simple viscous liquids (such as the theory underlying equation (3)). We will test whether this assumption holds for the case of the lightest nanoparticle, for which the hydrodynamic theory was shown to break down (Fig. 3). First, we calculate the relaxation of the fluid viscosity $\eta$ using equation (2) in the absence of an interface. Specifically, the relaxation of $\eta$ is characterized by the stress autocorrelation function (SAF). As shown in Fig. 4, the relaxation of SAF is significantly faster than the velocity relaxation of the lightest nanoparticle at a short timescale. Fitted to simple exponential decays, the solute relaxation time $\tau_1 \approx 1.7$ is indeed well separated from the viscosity relaxation time $\tau_2 \approx 0.07$. This separation of $\tau_1$ and $\tau_2$ explains why simulation results agree well with the theory for the Brownian motion of this nanoparticle in bulk liquid (see Fig. 2). On the other hand, it suggests that there is some other physics that is responsible for the discrepancy observed near a boundary and reported in Fig. 3. One possibility is the relaxation of the boundary, which means that the boundary condition may be time dependent. Although existence of such dynamic boundary has been hypothesized in literature[21–23], it has never been demonstrated before because of the challenges in calculating it theoretically and in measuring it experimentally. Consequently, a static boundary is assumed in most hydrodynamic theories.

Here we can test the hypothesis of boundary relaxation directly by calculating the friction coefficient $\bar{\eta}$ at the liquid–solid interface. To do that, we use our recently developed GK relation, which reads[17]

$$\bar{\eta} = \frac{1}{SkT(1-\alpha)} \int_0^\infty \sum_i \langle F(t)F(0)\rangle_i \, dt, \quad (4)$$

where $F$ is the friction force on the solvent particle $i$ exerted by the solid surface, $\alpha$ is the ensemble-averaged ratio between $F$ and the total force experienced by interfacial solvent particles and the sum is taken over all solvent molecules above surface area $S$. Similar to the viscosity, the boundary relaxation can be characterized by the friction force autocorrelation function (FAF) in equation (4). As shown in Fig. 4, the FAF initially follows the same relaxation trend as the SAF, but then it transitions to a slower exponential decay with a relaxation time $\tau_3 \approx 0.25$. The slow decay corresponds to a collective relaxation of structure of the first liquid layer near the surface. This result means that the onset of boundary conditions is not instantaneous, but instead it takes time to develop. More specifically, even in the case of atomically smooth and fully wetted liquid–solid interfaces with high friction at long timescales (a so-called static no-slip boundary), there can be a reduced friction and some slip present at shorter timescales. Consequently, a Brownian particle can experience a transition from partial-slip to no-slip boundary condition, which means that slip and the boundary condition are inherently dynamic properties. Such a dynamic picture of boundary condition is consistent with earlier observations of the dependence of slip length on shear rate[15,21,24] and frequency[17,25].

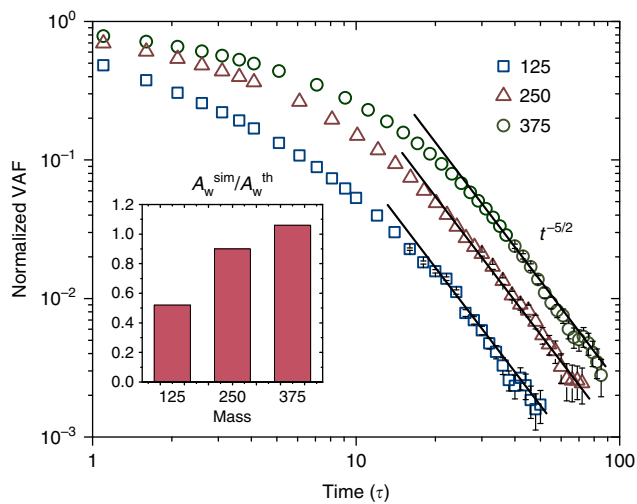

**Figure 3 | VAFs of nanoparticles with varying masses.** Log–log plot of the VAFs (normalized by its value at time zero) of nanoparticles ($R = 3\sigma$) with different masses near a static no-slip boundary. Solid black lines are added to guide the eye for the $t^{-5/2}$ asymptotic behaviour. The inset shows the ratio of amplitudes of the $t^{-5/2}$ long-time tail measured in simulations ($A_w^{sim}$) and predicted by theory ($A_w^{th}$). Data points represent averages over 5,000 independent simulations and error bars are obtained as s.d. from these averages. The error bars are not shown for VAF at short time scale, as they are smaller than the symbol sizes.

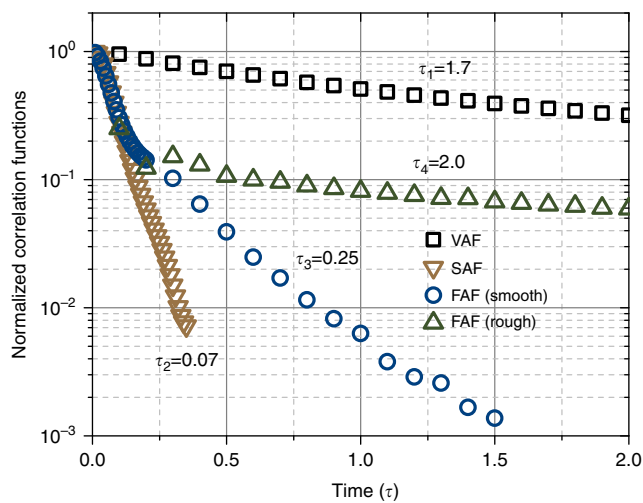

**Figure 4 | Correlation functions.** The normalized correlation functions that characterize the memories of different transport coefficients: the VAF for diffusion (squares), the SAF (stress autocorrelation function) for viscosity (down triangles) and the FAF (force autocorrelation function) for liquid–solid friction coefficient on atomically smooth surface (circles) and on rough surface (squares). Calculations are shown for $m_p = 125$.

The presence of reduced friction (or partial slip) at short timescales means that the liquid–solid coupling is weaker and in this regime the VAF of a Brownian particle decays slower (that is, has a higher amplitude) than expected for static no-slip boundary conditions. As the calculated zero-frequency diffusion constant (equal to the time integral of VAF) is the same as theoretically predicted for static no-slip boundary (see the inset in Fig. 2), the amplitude of long-time VAF has to be reduced to compensate for the short-time increase of the amplitude. It is noteworthy that the contribution of the long-time tail to the diffusivity is relatively small compared with the short-time VAF, and therefore a strong reduction at long-time scale is needed to compensate for the change in the short-time scale due to the boundary relaxation. In ref. 23, by assuming a simple Debye relaxation of the boundary condition, Felderhof found that boundary relaxation at $0.1\,\mu s$ can have a strong effect on the near-boundary VAF over the timescale of $10\,\mu s$. A large delay in the manifestation of a short-time boundary relaxation is also observed in our simulations. The suppression of VAF amplitude in the long-time regime (which we have shown to be due to the dynamic nature of boundary conditions) is consistent with the idea that hydrodynamic resonance shown in Fig. 1d would be suppressed near an interface. Specifically, as shown in ref. 26, PSD and VAF are related to each other via Fourier transform, and the smaller the magnitude of VAF, the smaller the amplitude of the resonance. As the magnitude of VAF is smaller near a boundary than in the bulk (see Fig. 2), it follows that the resonance will be suppressed for a particle near a boundary (Supplementary Note 2). This resonance suppression has been previously observed in experiments by Jannasch et al.[14] and it was attributed as a reduction of VAF amplitude with $t^{-3/2}$. Future work is needed to include a dynamic boundary discovered in our study in hydrodynamic theories, to determine whether the experiments of Jannasch et al.[14] agree with the $t^{-5/2}$ decay with reduced amplitude predicted by our study. On the other hand, our findings of the $t^{-5/2}$ decay in the long-time regime of VAF are consistent with optical tweezers experiments of Jeney et al.[13]

When considering relaxation of boundary conditions for engineering surfaces, one should take into account surface roughness, topology and even the possibility of forming air bubbles[23]. To shed light on this important issue, we introduce roughness by constructing a patterned surface with the pattern width of $8\sigma$ and height of $1\sigma$, and we investigate how this roughness has an impact on boundary relaxation. Surprisingly, we find that although the roughness is very small in our simulations, it has a dramatic effect on relaxation time. As shown in Fig. 4, the boundary (FAF) relaxation time on a rough surface is much slower ($\tau_4 \approx 2.0$) than on a smooth surface ($\tau_3 \approx 0.25$). As roughness of typical engineering surfaces can be often much higher than the one considered here, such roughness is expected to strongly affect nearby Brownian particles even with micrometre sizes. This sensitivity could be used as a potential

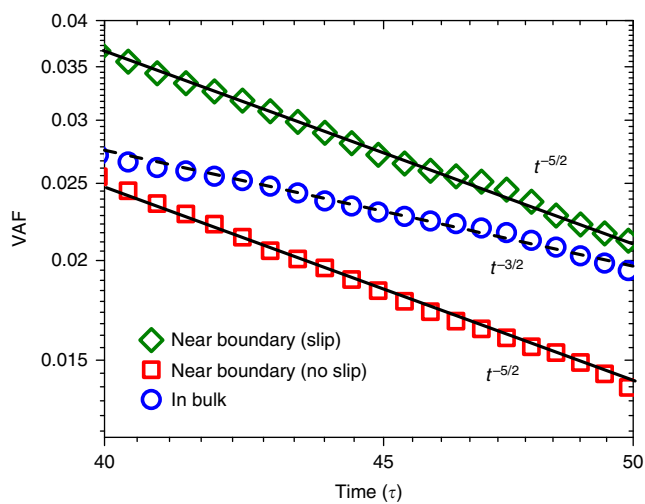

**Figure 5 | Effect of boundary conditions on VAF.** Long-time tails of VAF for a nanoparticle ($R = 3$, $M = 375$) near boundaries ($h = 5\sigma$) with different boundary conditions. Solid black line ($t^{-5/2}$ decay) and dashed black line ($t^{-3/2}$ decay) are added to guide the eye.

probe of the local properties of liquid–solid interfaces, such as wettability and slip length.

## Discussion

Most of the analysis so far was carried out for the case of static no-slip boundaries. Here we show the effect of a partially wetted surface on the nearby Brownian motion. As the partially wetted surface exhibits slip at long-time scales when the interfacial solvents are fully relaxed, we refer to it as a static slip boundary condition. We simulate such a boundary condition by using weak liquid–solid interaction strength $\varepsilon = 0.2$, which corresponds to a finite slip length $L_s$ of $23\sigma$. The VAF of a nanoparticle near such a partially wetted surface is shown in Fig. 5 and it is compared with VAFs of a particle near a wetted (no-slip) surface and in bulk. We find that although for a partially wetted surface the long-time tail of VAF still follows $t^{-5/2}$, the amplitude of VAF increases as compared with no-slip boundary conditions. In the limit of perfect slip ($L_s \to \infty$), some theories[10] predicted a $t^{-3/2}$ long-time tail. If these theories are correct, our results suggest that there could be a transition from the $t^{-5/2}$ to $t^{-3/2}$ as the slip length increases.

Our discovery of the long-time behaviour of VAF directly from molecular simulations addresses a long-standing debate in the field regarding the nature of the long-time tail near interfaces. We demonstrate that liquid–solid boundary conditions are inherently dynamic, and that relaxation of this boundary is important to account for in hydrodynamic theories and in models of micro-/nanoflow, even for perfectly smooth surfaces and even in the case of nominally no-slip boundaries. Although it had been known that interfaces disrupt the vortex generated by a nearby Brownian particle, we have shown that interfacial relaxation hinders the vortex backflow even further and this effect is expected to be significant for submicron solutes such as nanoparticles and biomolecules. The sensitivity of Brownian motion to the nearby L–S interface can enable advanced sensing applications, such as probing of local properties of L–S interfaces by monitoring the VAF of a single Brownian particle[6] or by measuring the two-point correlation function between two test particles[27].

## Methods

**Simulation system.** We construct a liquid box with dimensions of $64\sigma \times 64\sigma \times 128\sigma$ confined between two identical solid walls, where $\sigma$ is the unit of length in reduced Lennard Jones (LJ) units. Two-dimensional periodic boundary conditions are applied in the plane parallel to the interface. One spherical Brownian particle is immersed in the fluid near each interface and it is constrained by external harmonic potential along the direction normal to the interface. The average distance between the nanoparticle and the solid wall is kept at $5\sigma$. The nanoparticles interact with the solvent with a shifted LJ potential:

$$\phi(r) = 4\varepsilon \left[ \left( \frac{\sigma}{r-a} \right)^{12} - \left( \frac{\sigma}{r-a} \right)^{6} \right], \qquad (5)$$

where $a$ is the shifted distance of the potential, which defines the size of the nanoparticle. We choose $a = 2\sigma$ in our system, which yields a hydrodynamic radius of $R = 3\sigma$ suggested by literature[18]. The solvent–interface interaction and solvent–solvent interaction employ the same potential but with $a = 0$. The interaction strength is set to $1\varepsilon$ for solute–solvent and solvent–solvent interactions. For no-slip boundary condition, we choose the liquid–solid interaction strength to be $0.8\varepsilon$. A smaller value of $0.2\varepsilon$ is used for slip boundary condition in Fig. 5. As confinement effects may be more severe in the presence of large slip length, to avoid any such effects we use a box of size $48\sigma \times 48\sigma \times 200\sigma$ for the study of poorly wetted surfaces. A liquid cube with three-dimensional periodic boundary conditions is also built to study the unbounded Brownian motion as a reference. There are around $4 \times 10^5$ solvent particles in the bounded system and around $2 \times 10^5$ in the unbounded system. During the relaxation before data production, temperature is kept at 1.1 (in LJ units) and it is controlled with the Nose–Hoover thermostat. The thermostat is removed in the production stage. MD simulations are carried out using the LAMMPS software package (http://lammps.sandia.gov).

**Statistical sampling.** To have sufficient statistics in estimating long-time tails, we choose solvent density of 0.693, which has a low shear viscosity of 1.1 in reduced units. The time step is set to be $0.005\tau$, where $\tau = (m\sigma^2/\varepsilon)^{1/2}$ and $m$ is the mass of the liquid particle in LJ units. To obtain a good signal/noise ratio, 5,000 independent simulations with different starting configurations are carried out to obtain each near-boundary VAF curve. Each independent simulation consists of $10^5$ timesteps in the production stage. A cumulative $5 \times 10^8$ timesteps correspond to a total sampling time of $2.5 \times 10^6 \tau$ in reduced units or around 5 µs in real units for each VAF. Error bars are calculated as a s.d. from an average over 5,000 independent simulations.

## Acknowledgements

We gratefully acknowledge the financial support from the National Science Foundation (Grant CMMI-0747661) and the NSF-NSEC grant at the UW-Madison (Grant DMR-0832760). We thank the Center for High Throughput Computing (CHTC, UW-Madison) for the use of their computational facilities.

## Author contributions

K. H. and I. S. designed the simulations and wrote the manuscript. K. H. conceived the project, analyzed the data and performed the simulations.

## Additional information

**Supplementary Information** accompanies this paper at http://www.nature.com/naturecommunications

**Competing financial interests:** The authors declare no competing financial interests.

**Reprints and permission** information is available online at http://npg.nature.com/reprintsandpermissions/

**How to cite this article:** Huang, K. & Szlufarska, I. Effect of interfaces on the nearby Brownian motion. *Nat. Commun.* 6:8558 doi: 10.1038/ncomms9558 (2015).